\documentclass[11pt,a4paper]{article}
\usepackage[english]{babel}
\usepackage[utf8]{inputenc}
\usepackage[left=2.5cm,right=2.5cm,top=3cm,bottom=3cm]{geometry}
\usepackage{amsfonts}
\usepackage{amsmath}
\usepackage{booktabs}
\usepackage{caption}
\usepackage{amssymb}
\usepackage{mathrsfs}
\usepackage{amsfonts}
\usepackage{mathtools}
\usepackage{tensor}
\usepackage{graphicx}
\usepackage{appendix}
\usepackage{color}
\usepackage[dvipsnames]{xcolor}
\usepackage[affil-it]{authblk}
\usepackage[backend=bibtex,style=numeric-comp,maxnames=5,sorting=none,url=false,hyperref=true,eprint=false,doi=true,date=year]{biblatex}
\usepackage{hyperref}
\hypersetup{
	colorlinks=true,
	urlcolor=blue,
	linkcolor={black!40!black},
	citecolor={green!40!black}
}
\usepackage{cleveref}

\newcommand{\prn}[1]{\left(#1\right)}

\newcommand{\brkt}[1]{\left[#1\right]}

\newcommand{\defeq}{:=}
\newcommand{\ct}[2]{\tensor{{#1}}{#2}} 

\newcommand{\scri}{\mathscr{I}}
\newcommand{\dd}{\mathrm d}

\title{ Asymptotic Gravitational Radiation of Photon Rockets with $\Lambda$: Point, String, and Sheet Sources}
\author{Francisco Fernández-Álvarez\thanks{francisco.fernandez@ehu.eus}\ }
\author{José M. M. Senovilla\thanks{josemm.senovilla@ehu.eus}}
\affil[]{Departamento de Física\protect\\ Universidad del País Vasco UPV/EHU\protect\\ Apartado 644, 48080 Bilbao, Spain\protect\\\ }

\date{\today{}}

\begin{document}
\maketitle

\begin{abstract}
We study the existence of gravitational radiation at infinity for pure-radiation Robinson–Trautman metrics with one spacelike Killing vector and a cosmological constant $\Lambda$ of arbitrary sign. These metrics describe `photon rockets' with point, string, and sheet sources moving through spacetime while emitting null radiation. For $\Lambda\neq 0$, we use criteria for the existence of gravitational radiation based on the {\em asymptotic super-Poynting vector}. For point sources, we extend to $\Lambda\neq 0$ the known $\Lambda=0$ result that the Kinnersley rocket is the only rocket without gravitational radiation. However, for string and sheet sources, new possibilities without gravitational radiation arise. For $\Lambda>0$, we show that the vanishing of the canonical asymptotic super-Poynting vector---computed with respect to the unit normal to $\scri$---exactly characterizes the absence of radiation. On the other hand, for $\Lambda<0$, we show that the appropriate criterion is the vanishing of the components normal to $\scri$ of the asymptotic super-Poynting vectors associated with any unit vector tangent to $\scri$ or, equivalently, the proportionality between the Cotton–York tensor and the holographic stress tensor at $\scri$. In particular, we provide explicit examples of metrics for which the Cotton–York and holographic stress tensors commute, yet gravitational radiation is present because the two tensors are not proportional. We also determine the principal null directions (PNDs) of the rescaled Weyl tensor at $\scri$ for both signs of $\Lambda$ and relate their geometry relative to the normal to $\scri$ to the tensorial criteria for gravitational radiation.
\end{abstract}

\section{Introduction}
Robinson–Trautman (RT) spacetimes provide one of the simplest settings in which to study radiative gravitational fields in the presence of sources. In particular, the pure-radiation Robinson–Trautman solutions with axial symmetry include the well-known photon rockets \cite{Kinnersley:1969,Bonnor:1994,Bonnor:1996}, describing accelerated sources that emit null radiation and thereby generate a recoil. The Kinnersley photon rocket \cite{Kinnersley:1969} is the best-known example and plays a distinguished role as a solution with a non-trivial null-radiation field but without gravitational radiation. This was the subject of a controversy at the end of the last century \cite{Bonnor:1994}, and a detailed analysis using post-Minkowskian expansions \cite{Damour:1994} proved that the Kinnersley solution was an exceptional case, in which the anisotropic, purely dipolar photon emission generated a gravitational-wave amplitude that was exactly cancelled by the gravitational-wave amplitude due to the energy-momentum of the photons. Any other photon rocket, with non-dipolar anisotropic emission, should therefore have a net flux of gravitational radiation at infinity. This resolution was challenged by Bonnor \cite{Bonnor:1996}, who presented a particular family of photon rockets generalizing the Kinnersley solution and claimed that gravitational radiation was still absent. This turned out to be false, and several papers \cite{KramerGonna:1997,Cornish-Micklewright:1999,Cornish:2000} clearly demonstrated that {\em any} photon rocket in the RT family generates gravitational radiation reaching infinity {\em except} Kinnersley's rocket. All this was done in the realm of a vanishing cosmological constant, in which case the asymptotic structure at null infinity was well understood and the existence of gravitational radiation was characterized in terms of the news tensor (or function) \cite{Geroch1977,Bondi1962}.

Hence, the question of whether photon-rocket-like spacetimes radiate gravitationally at infinity becomes particularly interesting in the presence of a cosmological constant. Fortunately, the general metric describing such rockets with one spacelike Killing vector and an arbitrary cosmological constant $\Lambda$ was obtained by Podolský \cite{Podolsky:2008}, and the unphysical metric describing their conformal compactification was also presented there. Depending on the behaviour and topology of the two-dimensional wavefronts, the `rocket source' may be interpreted as point-like, string-like, or sheet-like. By using this family of solutions, we intend to ascertain whether the same conclusions as in the $\Lambda=0$ case remain valid, and which particular rockets do not produce gravitational waves at infinity.

When $\Lambda\neq0$, however, the notion of gravitational radiation at infinity has only been studied relatively recently \cite{Ashtekar2014,Fernandez-Alvarez-Senovilla2020b,Fernandez-Senovilla:2022b,FernandezSenovilla:2026}, and there is no widely accepted definition of a news tensor. In a series of papers, a fruitful framework for the characterization of gravitational radiation at infinity has been established, for arbitrary $\Lambda$, by using the asymptotic supermomentum and its spacelike part with respect to chosen observers, the {\em asymptotic super-Poynting vector} \cite{Fernandez-Alvarez-Senovilla2020b,Fernandez-Senovilla2020a,FernandezSenovilla:2026}. This is a vector field that measures the flux of tidal energy at infinity  with respect to the chosen observers. The framework provides covariant criteria for the absence of gravitational radiation at conformal infinity, formulated in terms of the electric and magnetic parts of the rescaled Weyl tensor when $\Lambda>0$, and in terms of the Cotton--York and holographic stress tensors when $\Lambda<0$. These criteria have been put to the test in several particular spacetimes \cite{Fernandez-AlvarezPodolskySenovilla:2024,ArenasCaimbelliDiazJiaRivera:2025,Diaz:2026,Fernandez-AlvarezSenovilla:2026b,Podolsky:2026}, with excellent results. In the present work, we apply our criteria to the complete family of pure-radiation RT metrics with one spacelike Killing vector and arbitrary cosmological constant, as presented by Podolský \cite{Podolsky:2008}. This provides a non-trivial test of the criteria in a broad family of exact radiative spacetimes and allows their predictions to be compared with the well-understood $\Lambda=0$ case.

Our results confirm that the asymptotic supermomentum and super-Poynting criteria \cite{FernandezSenovilla:2025}  work consistently in this setting and reproduce the expected physical picture. For point sources, the Kinnersley rocket remains the unique solution within this family without gravitational radiation, extending the known $\Lambda=0$ result to arbitrary non-zero $\Lambda$. For string and sheet sources, in contrast, additional non-radiative configurations arise. For $\Lambda>0$, the vanishing of the canonical asymptotic super-Poynting vector associated with the unit normal to $\scri$ exactly characterizes the absence of radiation. For $\Lambda<0$, where $\scri$ is timelike, the corresponding criterion is slightly more involved: the normal components of the  asymptotic super-Poynting vectors (or equivalently of the asymptotic supermomenta) \cite{FernandezSenovilla:2025} must vanish for every unit vector tangent to $\scri$. This is equivalent to the (functional) proportionality of the Cotton--York tensor of the conformal boundary and the holographic stress tensor \cite{FernandezSenovilla:2026}. Again, this criterion leads to the expected results.

The comparison between the two signs of the cosmological constant is particularly instructive. It shows that, for $\Lambda<0$, the vanishing of the commutator of the Cotton--York tensor with the holographic stress tensor, a condition advocated in \cite{CiambelliPasterskiTabor:2024} based on the analogy with the $\Lambda>0$ situation, need not be sufficient for the complete absence of radiation. In particular, for $\Lambda<0$ we provide explicit examples in which the Cotton--York and holographic stress tensors commute, but are not proportional, and gravitational radiation is nevertheless present.

Finally, we examine the same problem from the viewpoint of the principal null directions of the rescaled Weyl tensor. We determine these directions explicitly at $\scri$ for both signs of $\Lambda$ and analyse their geometry relative to the normal to conformal infinity. The resulting picture is fully consistent with the tensorial radiation criteria given in \cite{Fernandez-Alvarez-Senovilla2020b,Fernandez-Senovilla:2022b,FernandezSenovilla:2026}: the algebraic structure of the Weyl tensor, the causal character of $\scri$, and the asymptotic super-Poynting conditions lead to the same conclusions.

All in all, this paper provides an independent geometric confirmation of the radiation criteria based on the asymptotic  supermomentum and super-Poynting vectors and, in particular, shows the exceptional character of Kinnersley's rocket  ---and its versions with string and sheet sources presented herein--- even in the presence of a cosmological constant.

\section{Geometry and topology of the photon rockets}\label{sec:geom-top-rockets}
The metric analyzed in this paper belongs to the Robinson-Trautman family and, as given by Podolsky \cite{Podolsky:2008}, it can be written 
with coordinates $(u,r,x,\varphi)$ as
\begin{align}\label{eq:metric}
\dd s^2={}&
-\left(
-\frac12 G_{,xx}-\frac{2m(u)}r-\frac{\Lambda}{3}r^2
-r(bG)_{,x}-b^2Gr^2
\right)\dd u^2-2\dd u\,\dd r
\nonumber\\
&\quad+2br^2\,\dd u\,\dd x
+r^2\left(\frac{\dd x^2}{G}+G\,\dd\varphi^2\right),
\end{align}
where $m(u)$ and $G(x,u)$ are arbitrary functions---that we assume to be regular in $u$--- and $b(x,u)$ is then given by
\begin{equation}\label{eq:bx}
 b_{,x}=\left( \frac{1}{G}\right)_{,u} .
\end{equation}
Upon integration, this produces another arbitrary function of $u$. To keep the $(-,+,+,+)$ signature the function $G$ must be positive. The metric possesses a spacelike Killing vector $\partial_\varphi$.

The coordinate $u$ runs on the real line, $r\in \mathbb{R}^+$, and $\{x,\varphi\}$ are coordinates on the 2-surfaces defined by constant values of $u$ and $r$. Their ranges thus depend on the topology of these 2-surfaces, with the following possibilities:
\begin{enumerate}
\item\label{2zeros} If $G$ has two simple zeros\footnote{The case with double or higher zeros of $G$ will not be considered as they lead to curvature singularities.} $G(x_1,u)=G(x_2,u)=0$ with $G>0$ for all $x\in(x_1,x_2)$, then the topology is $\mathbb{S}^2$ with two connected components of an axis of symmetry, and the ranges are $x\in(x_1,x_2)$ and $\varphi \in (0,2\pi/C)$  for some constant $C$ that is defined in  \cref{sec:metric-topology-etc}  and rules the regularity of these axes.
\item\label{1zero} If $G$ has one simple zero $x_1$ with $G>0$ for all $x> x_1$, then there is an axis at $x=x_1$ and the topology is $\mathbb{R}^2$, with ranges $x\in(x_1,\infty)$ and $\varphi \in (0,2\pi/C)$.
\item\label{nozeros}  If $G$ has no zeros on $x$, then $x \in \mathbb{R}$ and there are two possibilities for $\varphi$: (i) $\varphi \in (0,2\pi)$ and the topology is cylindrical $\mathbb{R}\times \mathbb{S}^1$; and (ii) $\varphi \in \mathbb{R}$ and the topology is $\mathbb{R}^2$. An exceptional possibility arises if $G$ is periodic on $x$ with period $T$ and one identifies $x \leftrightarrow x+T$, in which case the topology is $\mathbb{T}^2$ for periodic $\varphi$.
\end{enumerate}

The metric \eqref{eq:metric} is a solution of the Einstein field equations with cosmological constant for a pure null radiation energy-momentum tensor
$$
T_{\mu\nu} = \frac{n^2(x,u)}{r^2} \ell_\mu \ell_\nu
$$
 the radiation traveling along the null vector field $\vec\ell := \partial_r$, which has orthogonal hypersurfaces $u=$ constant. The function $n^2$ is given explicitly by
 \begin{equation}\label{eq:n}
 n^2 = \left(-\frac{1}{8} GG''' +\frac{3}{2} m bG\right)'-m_{,u}
 \end{equation}
 where primes indicate derivatives with respect to $x$.
 The Gaussian curvature of the 2-surfaces defined by $u=$constant and $r=1$ is given by 
 \begin{equation}\label{gauss}
 {\cal K} = -\frac{1}{2} G'' \, .
 \end{equation}
 
 The interpretation of the function $m(u)$ depends on the topology of the 2-surfaces. The standard situation is case \ref{2zeros}, so that the spheres of constant $r$ can be interpreted as enclosing the trajectory of a particle that emits null radiation losing mass-energy, in which case $m(u)$ is the time dependent mass parameter. However, for the case \ref{nozeros} with cylindrical or toroidal topology, instead of a particle we would have a line, a string, emitting null radiation as it moves. In this situation the function $m(u)$ can be interpreted as a mass per unit length  parameter. Finally the cases \ref{1zero} and \ref{nozeros} with planar topology describe sheets emitting radiation and $m(u)$ represents the mass per unit area parameter.

 In what comes next, we shall use the following notation
\begin{equation}
A:=G_{,xxx}\ ,\qquad
K:=G G_{,uxx}+G_{,u}G_{,xx}-G_{,x}G_{,ux}\ ,
\qquad
L:=K-G^2bA\ .
\label{eq:AKL}
\end{equation}

\subsection{Generalization of the Kinnersley rockets}\label{sec:general-rocket}
As we will see presently, the conformal infinity $\scri$ of the above metrics will be locally conformally flat if and only if the function $K$ defined in \eqref{eq:AKL} and $A=G_{,xxx}$ both vanish,
\begin{equation}\label{eq:general-kinnersley}
A=0\ ,\quad K=0\ .
\end{equation}
From $G_{,xxx}=0$ one immediately gets
$$
G(x,u) = a(u) x^2 + c(u) x + e(u)
$$
and then from $K=0$ one easily derives
$$
c^2(u) -4 a(u) e(u) = k
$$
with a constant $k$. Therefore, $G$ can vanish somewhere if there exist real roots of the equation $G=0$, this is to say if
\begin{equation}\label{roots}
x_\pm (u) = \frac{-c(u)\pm \sqrt{k}}{2a(u)} 
\end{equation}
are real, which necessarily requires $k\geq 0$. 
Hence, the following possibilities arise:
\begin{itemize}
\item $k>0$ with $a(u)<0$. Then, there are two simple roots \eqref{roots} of $G$ and $G>0$ between them. To have axis regularity at one of the values, say $x_+$, one needs first of all that $x_+$ be independent of $u$, and then the regularity condition \eqref{cond} is automatically satisfied because from \eqref{roots} one has $2 a(u) x_+ +c(u) = \sqrt{k}$ which immediately implies  $G_{,xu} (x_+,u) = (2 a(u) x_+ +c(u))_{,u}=0$. Therefore, having another regular component of the axis at the other root also requires $G_{,xu} (x_+,u) = (2 a(u) x_- +c(u))_{,u}=0$ and then the three functions $a,c$ and $e$ are actually constants. The two roots are symmetrically placed with respect to the minimum of $G$ at $x_{min}= -c/(2 a)$ and, after some trivial redefinition of $x$, lead directly to the Kinnersley rocket metric with $\Lambda$ defined by
\begin{equation}
b=\alpha(u), \qquad G=1-x^2.
\label{eq:kinnersley}
\end{equation}
For $\Lambda =0$, the Kinnersley rocket \cite{Kinnersley:1969} was historically the first example of such `photon rockets', 

\item $k>0$ and $a(u)>0$. Now $G<0$ between the two simple roots \eqref{roots}, and thus one chooses $x\in (x_+,\infty)$. The second root $x_-(u)$ is outside that range and has no influence on the properties of the metric. The same reasoning as in the previous case leads to a constant $x_+$ and to $(2 a(u) x_+ +c(u))_{,u}=0$. Then, an elementary calculation proves
$$
G(x,u) = (x-x_+) \left[ a(u) (x-x_+) + \beta\right] , \hspace{1cm} \beta \in \mathbb{R}^+ 
$$
where $a(u)$ remains arbitrary as long as $a(u)>0$. The function $b$ in this case reads
$$
b(x,u) = \alpha(u) +\frac{a_{,u}}{a(u) \left[ a(u) (x-x_+) + \beta\right] } 
$$
with $\alpha(u)$ arbitrary.
\item $k<0$ so that there are no real roots of $G$, and then $a(u)>0$ to keep $G>0$ everywhere. Then $G$ can be written as
$$
G= a(u) (x- X(u))^2 -\frac{k}{4 a(u)} 
$$
with $2X(u) := -c(u)/a(u)$. In principle there remain two arbitrary functions of $u$, but $X(u)$ can be probably removed by redefining $x$. The function $b$ reads now
$$
b= \frac{2}{\sqrt{-k}} \arctan\frac{2 a(u) (x-X(u))}{\sqrt{-k}} +\alpha(u).
$$
\end{itemize}

In the first case (the Kinnersley rockets with $\Lambda$) the two components of the axis are regular ---see the Appendix--- and the topology of the 2-surfaces is $\mathbb{S}^2$, with $r^2$ measuring the area of these spheres except for a factor. The singularity at $r=0$ is then interpreted as a source emitting null radiation, with time-dependent mass parameter $m(u)$ and generally accelerating. Another outstanding family with these properties, but not satisfying $K=0=G_{,xxx}$ was studied by Bonnor \cite{Bonnor:1996} and generalized to have $\Lambda$ in \cite{Podolsky:2008}, and is given by the following function
\begin{equation}
G_B(x,u)=(1-x^2)\left[1+(1-x^2)H(x,u)\right],
\label{eq:GBonnor}
\end{equation}
where $H(x,u)$ is smooth at $x_1=-1$ and $x_2 =1$. The Kinnersley rocket \cite{Kinnersley:1969,Podolsky:2008} is the special subcase with $H=0$.

\subsection{Conformal infinity}

We are interested in studying the asymptotic structure of the metric and the content of gravitational radiation at infinity. For that, we first perform the change of variable
	\begin{equation}
		r \rightarrow \Omega=\frac{1}{r}\ .
	\end{equation}
Then, instead of working in physical spacetime $\prn{\hat{M},\hat{g}}$ with metric \eqref{eq:metric}, we consider the conformal compactification $\prn{M,g}$ with conformal metric $\ct{g}{_{\alpha\beta}}$
\begin{equation}
g_{\alpha\beta}=\Omega^2 \hat{g}_{\alpha\beta}\ .
\end{equation}
Explicitly, it reads
\begin{align}
\dd s^2={}&
\left[
\frac{\Lambda}{3}+Gb^2+\Omega(Gb)_{,x}
+\frac{\Omega^2}{2}G_{,xx}+2m\Omega^3
\right]\dd u^2+2\dd u\,\dd\Omega
\nonumber\\
&\quad+2b\,\dd u\,\dd x+\frac{\dd x^2}{G}+G\,\dd\varphi^2.
\label{eq:unphysical}
\end{align}
In this form, infinity is located at $\Omega=0$, which is the conformal boundary $\scri$ of $M$. This boundary is a differential manifold $\prn{\scri,h}$ with metric $\ct{h}{_{ab}}$. As is well known, depending on the sign of $\Lambda$ this metric is Riemannian ($\Lambda>0$), Lorentzian ($\Lambda<0$) or degenerate ($\Lambda=0$). From now on we will focus on the $\Lambda\neq 0$ cases.\\

Close to $\scri$, let us define
\begin{align}
\ct{N}{_{\alpha}}&\defeq \partial_{\alpha}\Omega ,\label{eq:normal}\\
\frac{1}{\sigma}\  N^2&\defeq -\ct{N}{^\mu}\ct{N}{_{\mu}}=
\frac{\Lambda}{3}+\Omega(Gb)_{,x}
+\frac{\Omega^2}{2}G_{,xx}+2m\Omega^3\ ,\quad \sigma\defeq \text{sign}(\Lambda)\label{eq:N}
\end{align}
This $\ct{N}{_{\alpha}}$ is the normal to $\scri$, and for $\Lambda\neq0$ one can define the unit normal
\begin{equation}
\ct{n}{_{\alpha}}\defeq \frac{1}{N}\ct{N}{_{\alpha}}\ ,\quad \ct{n}{^\alpha}=\frac{1}{N}\delta^\alpha_{u}- \frac{N}{\sigma}\delta^\alpha_{\Omega}-\frac{Gb}{N}\delta^\alpha_{x}\ .
\end{equation}
The coefficient $\sigma$ in \cref{eq:N} is necessary because, close to $\scri$, $\ct{N}{_{\alpha}}$ is timelike if $\Lambda>0$, and spacelike if $\Lambda<0$. Then, introduce the basis of one-forms close to $\scri$
\begin{equation}\label{eq:forms}
\omega^{0} \defeq -\frac{\sigma}{N}\dd\Omega\ ,\quad
\omega^{1} \defeq N\dd u + \frac{\sigma}{N} \dd \Omega\ ,\quad
\omega^{2} \defeq \frac{1}{\sqrt{G}}\prn{\dd x+ G b \dd u}\ ,\quad
\omega^{3} \defeq \sqrt{G}\dd \varphi
\end{equation}
such that
\begin{equation}
\dd s^2=\sigma\brkt{-\prn{\omega^0}^2+\prn{\omega^1}^2}+\prn{\omega^1}^2+\prn{\omega^{3}}^2\ .
\end{equation}
Later on, we will use also the basis of vector fields dual to \eqref{eq:forms},
\begin{equation}\label{eq:vectors}
e_0\defeq  \vec{n}\ ,\quad e_1\defeq \prn{\frac{1}{N}\partial_{u}-\frac{Gb}{N}\partial_{x}}\ ,\quad e_2\defeq \sqrt{G}\partial_{x}\ ,\quad e_3\defeq \frac{1}{\sqrt{G}}\partial_{\varphi}\ .
\end{equation}
The pull-back of $\omega^0$ to $\scri$ vanishes, whereas for the others one has
\begin{equation}\label{eq:forms-scri}
\overline{\omega}^{1} = \sqrt{\sigma\frac{\Lambda}{3}}\dd u ,\quad
\overline{\omega}^{2} = \frac{1}{\sqrt{G}}\prn{\dd x+ G b \dd u}\ ,\quad
\overline{\omega}^{3} = \sqrt{G}\dd \varphi\ ,
\end{equation}
and the dual basis on $\scri$ reads
\begin{equation}\label{eq:vectors-scri}
\overline{e}_1\defeq \prn{\frac{1}{N}\partial_{u}-\frac{Gb}{N}\partial_{x}}\ ,\quad \overline{e}_2\defeq \sqrt{G}\partial_{x}\ ,\quad \overline{e}_3\defeq \frac{1}{\sqrt{G}}\partial_{\varphi}\ .
\end{equation}
The induced intrinsic metric $\ct{h}{_{ab}}$ on $\scri$ reads,
\begin{equation}
h= \sigma \prn{\overline{\omega}^{1}}^2+\prn{\overline{\omega}^{2}}^2+\prn{\overline{\omega}^{3}}^2\ ,
\end{equation}
Thus, the sign of $\Lambda$, $\sigma$, determines the signature of the metric. In coordinates,
\begin{equation}
h=\frac{\Lambda}{3}\dd u^2+
G\left(b\,\dd u+\frac{\dd x}{G}\right)^2+G\,\dd\varphi^2\ .
\end{equation}
Observe that, implicitly, we are  identifying coordinates   $\prn{\overline{u},\overline{x},\overline{\varphi}}$ on $\scri$ with $\prn{{u},{x},{\varphi}}$, as this is the natural choice. 
Notice that the one form $b \dd u + G^{-1} \dd x$ is closed due precisely to \eqref{eq:bx}, therefore locally there exists a function $y(x,u)$ such that $\dd y=b \dd u + G^{-1} \dd x$ (equivalently $y_{,u}=b$ and $y_{,x}=1/G$). Using this new coordinate the metric at $\scri$ is diagonal in the chart $\{u,y,\varphi\}$
\begin{equation}
h=\frac{\Lambda}{3}\dd u^2+
G\left(\dd y^2+\,\dd\varphi^2\right)
\end{equation}
where now $G(y,u)$. Observe, however, that $y(x,u)$ will generically diverge at the zeros of $G$ and thus, for some geometrical analysis it is safer to use the original coordinates.

The topology of $\scri$ depends on the existence or not of zeros of $G$ at given values of $x$, as we discussed at the beginning of this section and further in \cref{sec:metric-topology-etc}. For the case \ref{2zeros} with two simple zeros of $G$, the topology of $\scri $ is $\mathbb{R}\times \mathbb{S}^2$; for the case \ref{1zero} with one simple zero the topology is $\mathbb{R}^3$, as it is also in case \ref{nozeros} when $G$ does not vanish and $\varphi\in \mathbb{R}$; finally, in case \ref{nozeros} and $\varphi \in (0,2\pi)$ the topology of $\scri$ is $\mathbb{R}^2 \times \mathbb{S}^1$,  or in the exceptional case $\mathbb{R}\times \mathbb{T}^2$. An important remark is in order here: while in the $\Lambda =0$ case the proper asymptotic properties imply that the topology of $\scri$ must be $\mathbb{R}\times \mathbb{S}^2$ \cite{Ashtekar2014,GerochHorowtiz1978,NewmanRPAC1989}, this is not the case when $\Lambda \neq 0$. For instance, the topology can be $\mathbb{S}^3$ (de Sitter and Taub-NUT-de Sitter), $\mathbb{R}\times \mathbb{S}^2$ (Kerr-de Sitter), or $\mathbb{R}^3$ (Kottler with non-compact preferred surfaces), etc.  \cite{MarsPaetzSenovilla2017}. Hence, there are no real restrictions on the topology of $\scri$ for cases with $\Lambda\neq 0$.
\\

\subsection{PNDs and Weyl scalars}
Close to $\scri$ we can define the following null tetrad using \eqref{eq:vectors}
\begin{equation}\label{eq:tetrad}
    \ell \defeq \frac{\sigma}{\sqrt{2}}\prn{-e_1+ e_0}\ ,\quad k \defeq \frac{1}{\sqrt{2}}\prn{e_1+e_0}\ ,\quad m\defeq \frac{1}{\sqrt{2}}\prn{e_2+i\, e_3}
\end{equation}
The vector field $\ell$ is the repeated principal null direction (PND). A direct computation of the Weyl scalars using this basis gives
\begin{align}
    \psi_2 &= -m \Omega \ ,\\
    \psi_3 &= \sqrt{\sigma \frac{\Lambda}{3}}\frac{q}{N}\Omega=\frac{\sqrt{G}A}{4 N}\, \Omega\ ,\\ 
    \psi_4 &=\frac{1}{4 N^2}\prn{\frac{\Lambda}{3}8 p\, \Omega - G \frac{\partial^4 G}{\partial x^4}\, \Omega^2}\ ,
\end{align}
where the functions
\begin{equation}\label{eq:pq}
q\defeq \sqrt{\sigma \frac{3}{\Lambda}}\frac{\sqrt{G}A}{4}\ ,\quad p\defeq \sigma\frac{3L}{4\Lambda G}\ ,
\end{equation}
will be used later on too, and $A$, $L$ are defined in \cref{eq:AKL}. It is clear, then, that the Weyl tensor vanishes at $\scri$,
\begin{equation}
\ct{C}{_{\alpha\beta\gamma}^{\delta}}\Big|_{\scri}=0\ .
\end{equation}
Hence, as usual, one can define the rescaled Weyl tensor
\begin{equation}
\ct{d}{_{\alpha\beta\gamma}^{\delta}}\defeq \frac{1}{\Omega}\ct{C}{_{\alpha\beta\gamma}^{\delta}}\ ,
\end{equation}
which is regular\footnote{See comments in \cref{sec:metric-topology-etc} on the regularity of the Weyl tensor; in particular, \cref{cond}.} and, in general, non-vanishing at $\scri$. The rescaled Weyl scalars at $\scri$ then read
\begin{equation}\label{eq:resc-scalars}
\phi_2\Big|_{\scri}=0\ ,\quad \phi_3\Big|_\scri=\frac{\sqrt{G}A}{4 N}\ ,\quad \phi_4 \Big|_{\scri} =\frac{8 p}{4 N^2}\frac{\Lambda}{3}\ .
\end{equation}

\section{Content of asymptotic gravitational radiation}
The characterization of gravitational radiation at $\scri$ using the formalism of \cite{Fernandez-Alvarez-Senovilla2022a,Fernandez-Senovilla:2022b,FernandezSenovilla:2026} is based on tidal ---traditionally called superenergy--- methods. In particular, the fundamental quantity is the \emph{asymptotic supermomentum}. The radiation condition is built upon it for the three possible cases $\Lambda>0$, $\Lambda<0$ and $\Lambda=0$. The theorems provide equivalent conditions that can be used to prove the existence of gravitational radiation in several equivalent ways. One of them uses the Cotton--York tensor $\ct{Y}{_{ab}}$ of the metric $h$ at $\scri$ and a traceless, symmetric, 3-dimensional tensor field $\ct{D}{_{ab}}$. In the case of $\Lambda>0$, the latter is the electric part of the rescaled Weyl tensor computed with respect to the  unit timelike normal; it is part of the covariant initial/final data at $\scri$ \cite{Friedrich1986a}. In the $\Lambda<0$ scenario, $\ct{D}{_{ab}}$ is not the electric part of the rescaled Weyl tensor, as the normal is spacelike, but plays a role in holography as the boundary holographic stress tensor\cite{Balasubramanian1999}; it is also closely related to the initial boundary problem, see \cite{Friedrich1995,FernandezSenovilla:2026} and references therein. Apart from using these tensors, an alternative (equivalent) analysis is to study the relative orientation of the PNDs with respect to the normal to $\scri$. We investigate both ways for the present metric.
\subsection{Cotton--York tensor and $D$}
Let us define the tensor fields at $\scri$
\begin{equation}
\ct{C}{_{\alpha\beta}}\defeq\frac{1}{2}\ct{n}{^\mu}\ct{n}{^{\nu}}\ct{\eta}{_{\alpha\mu\rho\sigma}}\ct{d}{^{\rho\sigma}_{\beta\nu}}\ ,\quad \ct{D}{_{\alpha\beta}}\defeq \ct{n}{^{\mu}}\ct{n}{^{\nu}}\ct{d}{_{\alpha\mu\beta\nu}}\ .
\end{equation}
Both tensors are orthogonal to $\ct{n}{_{\alpha}}$, and the pull-back of the tensor $\ct{C}{_{\alpha\beta}}$ to $\scri$ is
\begin{equation}
-\sqrt{\sigma\frac{\Lambda}{3}}\ct{C}{_{ab}}=\ct{Y}{_{ab}}\ ,
\end{equation}
where $\ct{Y}{_{ab}}$ is the Cotton--York tensor\footnote{Its vanishing is equivalent to local conformal flatness of the metric $h$.} of $\prn{h,\scri}$:
\begin{equation}\label{eq:CYdef}
Y_{ab}=\frac12\epsilon_a{}^{cd}C_{cdb}\ ,\quad C_{abc}=\nabla_cS_{ab}-\nabla_bS_{ac} ,\quad S_{ab}=R_{ab}-\frac14Rh_{ab}
\ .
\end{equation}
Here, $\ct{\eta}{_{\alpha\beta\gamma\delta}}$ is the space-time volume 4-form, and $\ct{\epsilon}{_{abc}}$ the induced volume element on $\scri$; we choose the orientation such that in an orthonormal frame $\ct{\epsilon}{_{123}}=1$\ . Using the computer algebra system Maxima, we calculate these tensors. Their form in the basis \cref{eq:forms-scri} reads
\begin{equation}\label{eq:D-C}
D_{\hat a\hat b}=
\begin{pmatrix}
-2m&-\sigma q&0\\
-\sigma q&\sigma m+p&0\\
0&0&\sigma m-p
\end{pmatrix}\ , \quad C_{\hat a\hat b}=
\begin{pmatrix}
0&0&-q\\
0&0&\sigma p\\
-q&\sigma p&0
\end{pmatrix}\ 
\end{equation}
where $p$ and $q$ are defined in \cref{eq:pq}, and recall that $\sigma=\text{sign}(\Lambda)$. Observe that conformal flatness of $h$ immediately implies the vanishing of $q$ and $p$, and therefore of $A$ and $K$ too ---see \cref{eq:AKL,eq:pq}--- i.e., the \emph{generalized} Kinnersley rockets \eqref{eq:general-kinnersley}  with $\Lambda$ and various topologies of \cref{sec:general-rocket}:
\begin{equation}
\ct{Y}{_{ab}}=0 \iff A=0=K\ \iff \text{generalized Kinnersley rockets of \cref{sec:general-rocket}.}
\end{equation}
 	\subsubsection*{Presence of radiation for $\Lambda>0$}
 	The criterion establishes that for $\Lambda>0$ there is no gravitational radiation at $\scri$ if and only if the asymptotic super-Poynting vector field $\ct{\overline{\mathcal{P}}}{^a}$ vanishes \cite{Fernandez-Alvarez-Senovilla2020b}. This is equivalent to the vanishing of the algebraic commutator of the magnetic $\ct{C}{_{ab}}$ and electric $\ct{D}{_{ab}}$ of the rescaled Weyl tensor. A computation using \eqref{eq:D-C} and the basis \eqref{eq:vectors-scri} yields
	 	\begin{equation}\label{eq:s-Poynting}
	 		\ct{\overline{\mathcal{P}}}{^{\hat{a}}}=\brkt{C,D}_{\hat{r}\hat{s}}\epsilon^{\hat{r}\hat{s}\hat{a}}=-2\prn{p^2+q^2}\delta^{\hat{a}}_{\hat{1}}+4q\prn{m-p}\delta^{\hat{a}}_{\hat{2}}\ .
	 	\end{equation}
	 Notice that $p^2+q^2=0 \iff p=0=q \iff A=0=K$. Consequently,
	 	\begin{equation}
	 	\brkt{C,D}=0 \iff \parbox{5cm}{No gravitational radiation at $\scri$ with $\Lambda>0$}\iff \parbox{4cm}{generalized Kinnersley rockets of \cref{sec:general-rocket}}
	 	\end{equation}
	 and, therefore, if and only if $(h,\scri)$ is conformally flat $\ct{Y}{_{ab}}=0$.
	\subsubsection*{Presence of radiation for $\Lambda<0$}
For the case with negative cosmological constant, the radiation criterion establishes that
	\begin{equation*}
		\text{There is no gravitational radiation }\iff \ct{\overline{\mathcal{P}}}{^\alpha}\prn{u}\ct{n}{_{\alpha}} = 0 \quad \forall \ct{u}{^\alpha}\ ,
		\end{equation*}
	where $\ct{\overline{\mathcal{P}}}{^\alpha}\prn{u}$ is the asymptotic super-Poynting vector field for a timelike unit observer $\ct{u}{^\alpha}$ tangent to $\scri$ ($\ct{u}{^\alpha}\ct{n}{_{\alpha}}=0$). The theorems of \cite{FernandezSenovilla:2026}
	provide an equivalent condition in terms of $\ct{C}{_{ab}}$ and $\ct{D}{_{ab}}$: radiation is absent if and only if they are proportional,
		\begin{equation}\label{eq:crit-neg-cc}
			\beta \ct{D}{_{ab}}=\gamma \ct{C}{_{ab}} \iff \text{No gravitational radiation at $\scri$ with $\Lambda<0$}\ ,
		\end{equation}
	where $\beta$ and $\gamma$ are functions. Inspecting \cref{eq:D-C}, one realises that the condition is equivalent to $q=0=p$, thus one reaches the same conclusion as in the case $\Lambda>0$:
		\begin{equation}\label{eq:result-neg-cc}
		 	\beta\ct{D}{_{ab}}=\gamma \ct{C}{_{ab}}  \iff\parbox{5cm}{No gravitational radiation at $\scri$ with $\Lambda>0$}\iff \parbox{4cm}{generalized Kinnersley rockets of \cref{sec:general-rocket}}
	 	\end{equation}
	 and, again, this happens for $A=0=K$, i.e., the conformally flat case $\beta=0$. It is important to remark that condition \eqref{eq:crit-neg-cc} is not equivalent to the vanishing of the commutator; in fact, it is stronger (in the sense that it contains fewer non-radiating cases). In \cite{CiambelliPasterskiTabor:2024} the authors used the commutator, instead of the proportionality condition. Unfortunately, the commutator does not always guarantee the correct result, as it was largely justified in \cite{FernandezSenovilla:2026}. In fact, we can use the present metric as an example to illustrate this. If one computes the commutator in the basis \cref{eq:vectors-scri} the result reads ---see \cref{eq:pq}---
	 \begin{equation}\label{eq:comm-neg-cc}
	 	\brkt{C,D}_{\hat{r}\hat{s}}\epsilon^{\hat{r}\hat{s}\hat{a}}= 2\prn{2 p^2-q^2}\delta^{\hat{a}}_{1}+2q\prn{2p+3m}\delta^{\hat{a}}_{2}\ .
	 \end{equation}
Observe the difference between \cref{eq:comm-neg-cc} and \cref{eq:s-Poynting}: the term in $\delta^{\hat{a}}_{1}$ was a sum of squares in the latter, whereas in the present case it is  $2 p^2-q^2$. This prevents \eqref{eq:comm-neg-cc} from giving the right answer. Indeed, one has
	\begin{equation}\label{eq:comm-neg-cond}
		\brkt{C,D}_{rs}=0 \iff 
		\begin{dcases}
		6\frac{L^2}{G^2}+\Lambda GA^2=0.\\
	\frac{\sqrt{G}A}{4}\prn{-\frac{L}{2\Lambda G}+m}=0,
		\end{dcases}
	\end{equation}
The solutions to this condition can be classified into two branches
	\begin{enumerate}
	\item $q=0$. This yields, $A=0=K$, that is the conformally flat case. Thus, these are included in the radiation criterion as non radiative \cref{eq:result-neg-cc}.
	\item  $q\neq 0$. This is a different branch for which necessarily
		\begin{align}
24\Lambda m^2 + G A^2 &=0\ ,\label{eq:branch-q}\\
		-\frac{L}{2\Lambda G}+m&=0	 \label{eq:branck-q-2}\ .
		\end{align}
	The radiation criterion determines that these are \emph{radiating} solutions, as the proportionality condition is not fulfilled. Observe that \cref{eq:branch-q} implies that $G$ cannot have zeros, meaning that these solutions do not possess axial symmetry ---see \cref{sec:geom-top-rockets,sec:metric-topology-etc} for more details on the topology of these cases. 
	\end{enumerate}
\subsection{Characterization using PNDs and Weyl scalars}
The theorems of \cite{FernandezSenovilla:2026,Fernandez-Senovilla:2022b} state that the radiation criterion is equivalent to some geometrical conditions on the PNDs of the rescaled Weyl tensor at $\scri$. In the case of a type-II/D metric they simplify, reading with $\Lambda<0$
	\begin{equation}\label{eq:cond-neg-cc}
		  \parbox{3cm}{No gravitational radiation at $\scri$ with $\Lambda<0$ for type II/D} \iff
		  \begin{dcases*}
		 \parbox{10cm}{One of the following holds:
		 \begin{itemize}
		 \item All principal null directions of the rescaled Weyl tensor are tangent to $\scri$.
		 \item The normal is coplanar with two principal null directions of the same multiplicity and the remaining PNDs (if any) are tangent to $\scri$.
		 \end{itemize}}
		 \end{dcases*}
	\end{equation}
and for type II/D with $\Lambda>0$
		\begin{equation}
			  \parbox{3cm}{No gravitational radiation at $\scri$ with $\Lambda>0$ for type II/D} \iff
			 \parbox{8cm}{
			  The normal is coplanar with two repeated principal null directions.}
		\end{equation}
Since we are going to study the type-D subcases (both radiative and non-radiative), we compute the other 2 PNDs at $\scri$ other than the repeated one  $\ell$. For that, we do a null rotation of \cref{eq:tetrad}, keeping $\ell$. This yields a new null direction
	\begin{equation}
		k'=k+cm+\bar{c}\bar{m}+c\bar{c}\ell\ ,
	\end{equation}
where $c$ is the complex function giving the null rotation ---see, for instance, \cite{Stewart1991}. Alternatively, in terms of \cref{eq:vectors} one can write this null direction as:
	\begin{equation}
	k'=\frac{1}{\sqrt{2}}\brkt{\prn{1+\sigma c\bar{c}} e_0+\prn{1-\sigma c\bar{c}} e_1+\prn{c+\bar{c}}e_2+i(c-\bar{c})e_3}\ ,
	\end{equation}
 The condition for $k'$ to be a PND at $\scri$ is simply
	\begin{equation}
		\phi_4 +4c\phi_3+6c^2\phi_2=0\ .
	\end{equation}
Substituting \cref{eq:resc-scalars},
	\begin{equation}
	p+2qc-3mc\bar{c}=0
	\end{equation}
	and solves to
	\begin{equation}
	c_{\pm}=\frac{q\pm\sqrt{q^2+3mp}}{3m}\ .
	\end{equation}
	Recall that the functions $p$ and $q$ \eqref{eq:pq} are different depending on the sign of $\Lambda$. Thus the other two PND read:
		\begin{equation}\label{eq:other-PND}
			k'_{\pm}=k+c_{\pm}m+\bar{c_\pm}\bar{m}+c_\pm\bar{c_\pm}\ell\ .
		\end{equation}	
	The condition that $k_{-}=k_{+}=k'$, i.e., that there is a second repeated PND (and thus, a type-D rescaled Weyl tensor at $\scri$)	is simply $c_{+}=c_{-}=c$, or
		\begin{equation}
		q^2+3mp=0\ .
		\end{equation}
	 If we use \cref{eq:pq,eq:AKL} to substitute for $q$ and $p$, the condition reads,
			\begin{equation}\label{eq:type-D-eq}
					 GA^2+12m\prn{\frac{K}{G}-12mbAG}=0\ .
			\end{equation}
	Observe that this equation does not depend on $\sigma$, the sign of $\Lambda$, and allows $G$ to have zeros, and that if it holds, it makes the unique double solution $c$ real. Therefore, there are two branches of \textbf{type-D solutions},
		\begin{enumerate}
		\item $A=0$. Then $K=0$ too, and we are in the conformally flat case corresponding to the \emph{generalized} Kinnersley rockets of \cref{sec:general-rocket}. Also, since $c=0$, one has that $k=k'$ is the repeated PND. The two repeated PNDs read
		\begin{equation}
			\sqrt{2}\ell=\sigma \prn{ e_0 -e_1}\ ,\quad\sqrt{2}k=e_0+e_1\ .
		\end{equation}
		Thus, they are coplanar with the normal $n=e_0$.
		\item $A\neq 0$. This is again a type-D case, but this time the PND $k$ and $\ell$ are not coplanar with the normal $n=e_0$, since
		\begin{equation}
			k=\frac{1}{\sqrt{2}}\brkt{\prn{1+\sigma c^2} e_0+\prn{1-\sigma c^2} e_1+2ce_2}\ ,
			\end{equation}
		where we have used that $c$ in this case is real. The $k$ only becomes coplanar with $\ell$ and $n$ when $c=0$, but then one falls in case 1. We comment more on this branch in what comes next.
		\end{enumerate}
	\subsubsection*{Presence of radiation for $\Lambda>0$}
	In this case it is clear that the no radiation condition can only hold in case 1 of type-D. All Petrov type-D solutions in case 2 are radiating at $\scri$. 
	In addition, one can check this using the criterion \cite{Fernandez-Alvarez-Senovilla2020b} in terms of the rescaled Weyl scalars:
		 	\begin{equation}
		 			  \parbox{3cm}{No gravitational radiation at $\scri$ with $\Lambda>0$} \iff \begin{dcases}
		 			  2\prn{\phi_1\bar{\phi}_1-\phi_3\bar{\phi}_3}+\phi_0\bar{\phi}_0-\phi_4\bar{\phi}_4=0\ ,\\
		 			  \phi_3\bar{\phi}_4+\phi_1\bar{\phi}_0-3(\phi_1+\phi_3)\phi_2=0\ .
		 			  \end{dcases}
		 		\end{equation}
	Looking at \cref{eq:resc-scalars} we find that there is no radiation if and only if
		\begin{equation}
			K=0=L\ ,
		\end{equation}
	thus, same result as in our previous analysis.
	\subsubsection*{Presence of radiation for $\Lambda<0$}
	It is clear that the first of \cref{eq:cond-neg-cc} cannot happen, as $\ell$ \eqref{eq:tetrad} always has a component in $n$. Thus, we are to deal with the second possibility:
	\begin{itemize}
	\item For type D, i.e., when \cref{eq:type-D-eq} holds, $n$ is coplanar with the 2 repeated PNDs if and only if one has $c=0$, i.e., case 1 above. This is the only case without radiation for type D, i.e., the \emph{generalized} Kinnersley rockets. The other case 2 of type-D metrics are always radiating.
	\item Type II is always radiating, as the repeated PND will never become tangent to $\scri$.
	\end{itemize}
	Once again, the criterion \cite{FernandezSenovilla:2026} presents a formulation in terms of rescaled Weyl scalars,
		\begin{equation}
			\alpha\bar{\phi}_4=-\bar{\alpha}\phi_{0}\ ,\quad 	\alpha\bar{\phi}_1=-\bar{\alpha}\phi_{3}\ ,\alpha\bar{\phi}_2=-\bar{\alpha}\phi_{2}\ ,
		\end{equation}
	where $\alpha=\beta+i \gamma$ and $\beta$ and $\gamma$ are the functions appearing in \cref{eq:crit-neg-cc}. Inspection of \cref{eq:resc-scalars}	yields
	\begin{equation}
	\beta=0\ ,
	\end{equation}
		which holds if and only if $\prn{\scri,h}$ is conformally flat, i.e., the rocket solutions of \cref{eq:general-kinnersley}, again.
	\subsubsection*{Type-D radiating solutions}
	We have seen that both for $\Lambda<0$ and $\Lambda>0$ the type-D solutions of case 2 ($A\neq 0$) are radiating at $\scri$. In principle, they admit zeros $x_0$ of $G(x,u)$, resulting in different possible topologies ---see \cref{sec:geom-top-rockets}--- containing radiation at $\scri$. A particular subcase is those solutions having $K=0$, hence $A=12mb$. In general, if $K\neq 0$, one has to require
		\begin{equation}
		\lim_{x\rightarrow x_0}\frac{G}{K}=\text{finite}\ .
		\end{equation}
	Let us have a look at the content of gravitational waves at $\scri$:
		\begin{itemize}
		\item If $\Lambda>0$. These are just type-D solutions with the PNDs not coplanar with the normal. Other examples of this kind of radiation are the ones produced by the generalized type-D accelerated black holes \cite{Fernandez-AlvarezPodolskySenovilla:2024}.
		\item  If $\Lambda<0$. As we have seen, the  proportionality of $C$ and $D$ is not fulfilled, which definitely characterizes the solutions as radiative. Regarding their commutator, substituting \cref{eq:type-D-eq} for $L$ in \cref{eq:branck-q-2}, it follows that these solutions have a non-vanishing commutator too.
		\end{itemize}
	\subsubsection*{Presence of radiation for $\Lambda=0$}
	Even though we are focusing on the case $\Lambda\neq 0$, we can recover easily the results on $\Lambda=0$ that found absence of gravitational radiation for the Kinnersley rocket \cite{Damour:1994,KramerGonna:1997,Cornish:2000} and extend them to the generalized solutions considered in \cref{sec:general-rocket}. We can do the limit \cite{Fernandez-Senovilla:2022b} from $\Lambda>0$ ($\sigma=1$) to $\Lambda=0$ by noting that the tetrad $\prn{M,K,m}$, with
		\begin{align}
		    M \defeq& \lim_{\Lambda\rightarrow 0}\frac{1}{N} \ell= -\frac{1}{\sqrt{2}}\partial_\Omega\ ,\\
		    K \defeq& \lim_{\Lambda\rightarrow 0}N k=\sqrt{2}\prn{\partial_{u}-bG\partial_{x}}\ 
		\end{align}
	and $m$ as in \cref{eq:tetrad}, is well defined at infinity when $\Lambda=0$. In fact, $M$ is still the repeated PND, and $K$ is proportional to the null normal $N$ at $\scri$, $K=\sqrt{2}N$. The rescaled Weyl scalars now read at $\scri$
		\begin{equation}
		 		\phi_4=\frac{L}{2G}\ ,\quad \phi_{3}=-\frac{A}{4}\sqrt{G}\ ,\quad 	\phi_2=-m\ .
		\end{equation}
	It is straightforward to compute the asymptotic radiant supermomentum \cite{Fernandez-Senovilla2020a} using the expressions in terms of rescaled Weyl scalars of appendix D of \cite{Fernandez-Senovilla:2022b},
		\begin{equation}
		\mathcal{Q}^\alpha=\frac{L^2}{G^2}M^\alpha+\frac{A^2}{4}G K^\alpha+\frac{AL}{2\sqrt{G}}\prn{m^\alpha+\bar{m}^\alpha}\ .
		\end{equation}
	The criterion establishes that there is no gravitational radiation if and only if $\mathcal{Q}=0$. Therefore
	\begin{equation}
	 \parbox{5cm}{No gravitational radiation at $\scri$ with $\Lambda=0$}\iff \parbox{4cm}{generalized Kinnersley rockets of \cref{sec:general-rocket}}
	\end{equation}

\section*{Acknowledgments}
JMMS is grateful to Prof. Ji\v{r}\'{i} Podolský for bringing the subject of photon rockets to his attention and providing reference \cite{Podolsky:2008}. 
FFA was partly supported by a contract of the project PID2021-123226NB-I00, financed by the Spanish MICIU/AEI and FEDER (European Union). 

\appendix
 \section{Axes of symmetry and singularities}\label{sec:metric-topology-etc}

Concerning singularities of the metric, if we are in case \ref{nozeros} { as defined in the main text (\cref{sec:geom-top-rockets}) and $G$ is $C^3$, say, then there are no divergences in the function $n^2$ of \eqref{eq:n} for finite values of $x$. The only divergences that may occur, leaving aside $r=0$, would then appear at $x\rightarrow \pm \infty$. They are inevitable for asymptotic exponential growth of $G$, but if $G\sim g(u) |x|^p$ as $|x|\rightarrow \infty$, then there are no problems for $p<1$, and for $1<p\leq 2$ the divergence can be avoided if $\lim_{|x|\rightarrow \infty} b =0$. This is only possible, combining both asymptotics at $\pm \infty$, if
$$
\frac{d}{du} \left(\int_{-\infty}^\infty \frac{1}{G} dx \right) =0.
$$
 
For the other cases \ref{1zero} and \ref{2zeros}, $G$ has simple zeros at some fixed values of $x$, e.g. at $x=x_1$. Then, even though $b$ diverges there, the term $(bG)'$ is finite if 
\begin{equation}\label{cond}
G_{,xu}(x_1,u) = 0.
\end{equation}
One can further check that this also ensures the regularity of the Weyl tensor.
We are going to see that \eqref{cond}  is precisely the condition ensuring the regularity of $x=x_1$ at an axis of symmetry. 

 Assume that $G$ vanishes at $x=x_1(u)$: $G(x_1(u),u)=0$. In principle, this might define a time-dependent axis of symmetry. To check its possible regularity we use $|\partial_\varphi|^2 = r^2 G$ and compute
 $$
 \lim_{x\rightarrow x_1(u)} |\dd (r \sqrt{G})|^2 = \frac{1}{4} G'^2(x_1(u),u) -r G_{,u}(x_1(u),u) .
 $$
 which must be a constant \cite{MarsSenovilla1993}. But this necessarily implies that $G_{,u}(x_1(u),u)=0$ which, together with $G(x_1(u),u)=0$ readily leads to \footnote{Unless $G'(x_1(u),u) =0$, but this is a double zero of $G$ at $x_1$, and they can be checked to signal a curvature singularity at $x_1$.}.
 $$
 x_{1,u}=0
 $$
 so that $x_1$ is constant. Thus, the condition reduces to
 $$
 \frac{1}{4} G'^2(x_1,u) = C^2
 $$
 with a {\em constant} $C$, as otherwise there would be a time-dependent deficit/excess angle that cannot be removed. Thus the condition for regularity of the axis at $x=x_1$ is precisely \eqref{cond}
 and the range of the coordinate $\varphi'=C \varphi $ is the required $2\pi$.
 
 In the case that there are two components of the axis, $x_1$ and $x_2>x_1$, the regularity at one of the axes, say $x=x_1$, can always be achieved by the previous choice of $C$, but the regularity on the second one then requires
 $$
 |G'(x_1,u)| =|G'(x_2,u)|
 $$
 which does not need to be satisfied in general. If this condition does not hold, there is a deficit/excess angle at the axis $x=x_2$.

As a final comment, in case \ref{nozeros} with $G> 0$ everywhere, the 2-surfaces cannot keep a positive Gaussian curvature everywhere due to \eqref{gauss}, because a concave curve on the whole real line must go to $-\infty$ at least on one side. However, for the case \ref{1zero}, a positive Gaussian curvature is certainly possible. An explicit example (for a complete surface) is
$$
G = 2C\ln(1+x) 
$$
for a constant $C$. This $G$ vanishes at $x=0$ where there is a regular axis with $C$ the constant that defines the length of $\varphi$ as explained before: $G'(0,u) =2C$. The Gaussian curvature is simply ${\cal K} = (1+x)^{-2}$. Observe that, for this case,
$$
\int {\cal K} dA = 2\pi \int_0^\infty {\cal K} =2\pi
$$
saturating the Cohn-Vossen \cite{ShiohamaShioyaTanaka2003} inequality for complete non-compact Riemannian surfaces. The function $b=\alpha (u) $ for the solution with this $G$.

\printbibliography

\end{document}